\documentclass[10pt, paper=a4, UKenglish]{article}
\usepackage{graphicx}
\def\Title#1{\begin{center} {\Large #1 } \end{center}}
\def\Author#1{\begin{center}{ \sc #1} \end{center}}
\def\Address#1{\begin{center}{ \it #1} \end{center}}

\newcommand\pubblock{\rightline{\begin{tabular}{l} Proceedings of the CTD/WIT 2019\\ \pubnumber\\
			\pubdate  \end{tabular}}}

\newenvironment{Abstract}{\begin{quotation} \begin{center} 
			\large ABSTRACT \end{center}\bigskip 
		\begin{center}\begin{large}}{\end{large}\end{center} \end{quotation}}

\newenvironment{Presented}{\begin{quotation} \begin{center} 
			PRESENTED AT\end{center}\bigskip 
		\begin{center}\begin{large}}{\end{large}\end{center} \end{quotation}}

\def\beq{\begin{equation}}
\def\eeq#1{\label{#1}\end{equation}}
\def\eeqn{\end{equation}}

\def\beqa{\begin{eqnarray}}
\def\eeqa#1{\label{#1}\end{eqnarray}}
\def\eeqan{\end{eqnarray}}

\let\bar=\overbar

\def\Dslash{\not{\hbox{\kern-4pt $D$}}}
\def\dslash{\not{\hbox{\kern-2pt $\del$}}}

\def\msb{{\bar{\ssstyle M \kern -1pt S}}}

\usepackage{color}
\usepackage{lineno}
\usepackage{subfig}
\usepackage{hyperref}

\newcommand\pubnumber{PROC-CTD19-024}

\newcommand\pubdate{\today}

\def\affiliation{
	on behalf of the LHCb collaboration, \\
	IFIC \\
	CSIC-Universitat de Val\`encia, Spain}

\newcommand{\conference}{Connecting the Dots and Workshop on Intelligent Trackers (CTD/WIT 2019)\\
	Instituto de F\'isica Corpuscular (IFIC), Valencia, Spain\\ 
	April 2-5, 2019}

\usepackage{fancyhdr}
\definecolor{mygrey}{RGB}{105,105,105}
\begin{document}
	
	
	\large
	\begin{titlepage}
		\pubblock
		
		\vfill
		\Title{Tracking performance for long-lived particles at LHCb}
		\vfill
		
		\Author{Luis Miguel Garcia, Louis Henry, Brij Jashal, Arantza Oyanguren}
		
		\Address{\affiliation}
		\vfill
		
		\begin{Abstract}
			The LHCb experiment is dedicated to the study of the $c-$ and $b-$hadron decays, including long-lived particles such as $K_s$ and strange baryons ($\Lambda^0$, $\Xi^-$, etc... ). These kind of particles are difficult to reconstruct by the LHCb tracking system since they escape detection in the first tracker. A new method to evaluate the performance of the different tracking algorithms for long-lived particles using real data samples has been developed. Special emphasis is laid on particles hitting only part of the tracking system of the new LHCb upgrade detector.
		\end{Abstract}
		
		\vfill
		
		\begin{Presented}
			\conference
		\end{Presented}
		\vfill
	\end{titlepage}
	\def\thefootnote{\fnsymbol{footnote}}
	\setcounter{footnote}{0}
	%
	
	\normalsize 
	
	
	\section{Introduction}
	
	\subsection{Importance of Long-lived particles}
	
	Long-lived particles (LLPs) produced in proton-proton collisions at LHC are key for both the study of the Standard Model (SM) of particle physics and to search for new physics beyond it. Many interesting decay modes involve strange particles with large lifetimes such as $K_s$ or $\Lambda^0$. Exotic LLP are also predicted in many new theoretical models. Examples of these are rare radiative decays of $\Lambda^0_b$ baryons~\cite{Lambdab} and decays of new Higgs-like bosons into $\pi_V$~\cite{LLP1}.
	
	The selection and reconstruction of LLPs at the LHCb experiment is a challenge. These particles can decay far from the primary interaction vertex and are hard to identify. Monitoring the performance of the present tracking algorithms for LLPs is key for the understanding of many physics analyses, and to be able to develop new techniques to improve the algorithms. In the following, after introducing the LHCb detector, a novel method to determine the performance of the dedicated tracking algorithms for LLPs at LHCb is explained. 
	

	

	\subsection{LHCb detector}
\begin{figure}[htbp]
		\centering
		\includegraphics[width=0.48\linewidth]{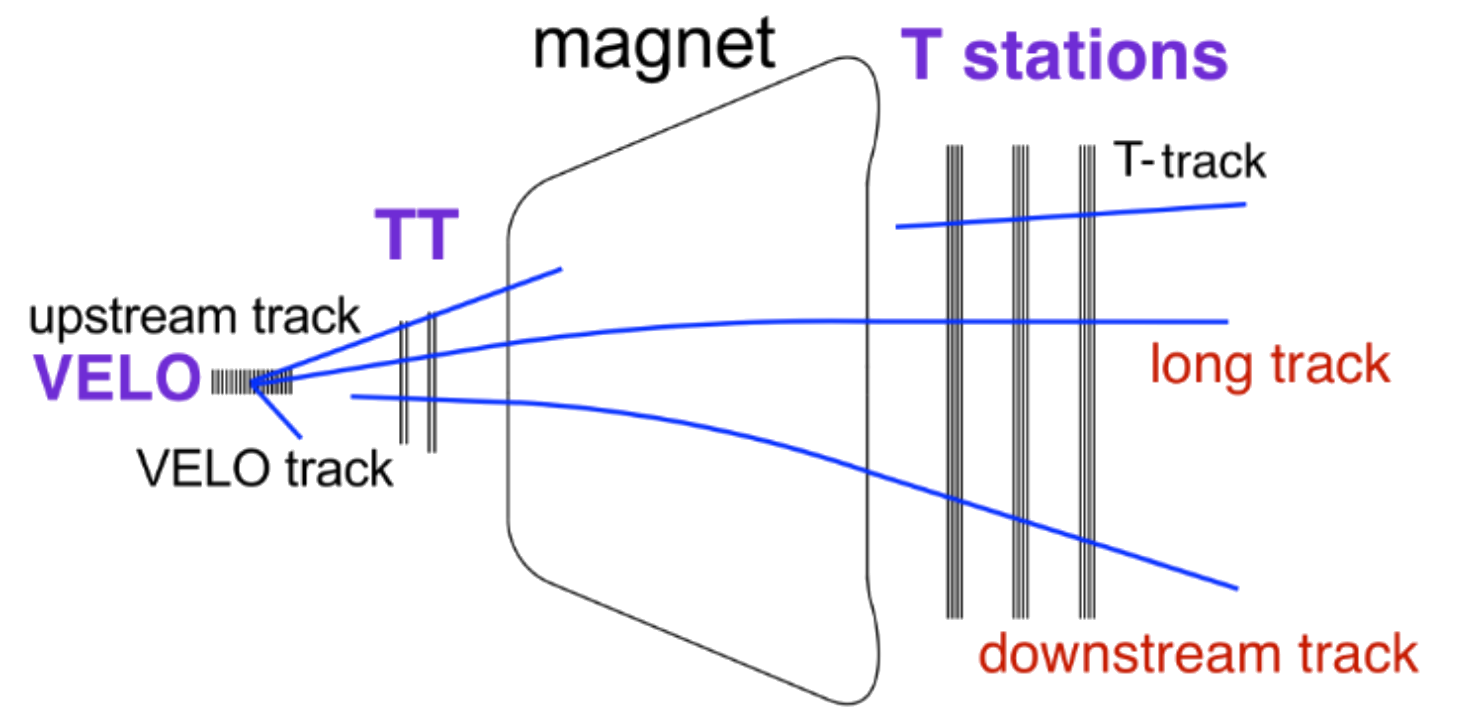}
		\caption{Tracks types and tracking system at LHCb.}
		\label{fig:trackstypes}
	\end{figure}

	The LHCb tracking system for charged particles consists of three subsystem: VELO, TT and T-stations, and a magnet to bend particles in order to estimate their momentum~\cite{LHCb}. The particles will have a different track type depending on the subsystem used to reconstruct their track as is shown in Figure~\ref{fig:trackstypes}.
	
	Out of all these categories, the relevant track types for this are:
	\begin{itemize}
		\item \textbf{Long tracks}, reconstructed with hits from at least the VELO and T-stations. The VELO subdetector is the closest to the interaction point and  the most densely instrumented providing high precision tracking. Due to this, the long tracks have the best spatial and momentum resolution among all track types and therefore are used in most of the analyses at LHCb.
		\item \textbf{Downstream tracks}, reconstructed with hits from the TT and T-stations. They are mainly the decay products of long-lived particles and have lower resolution and efficiency, since the information from the VELO subdetector is not available.
		
	\end{itemize}


	Despite their lower performance, the inclusion of downstream tracks could benefit significantly analyses involving long-lived particles. Decay modes suffering from low statistics, such as rare b-baryon decays, are particularly affected. An example would be the recent measurement of the decay $\Lambda_{b}^{0} \to \Lambda^0 \gamma$~\cite{Lambdab} which uses only long tracks. If downstream tracks were also used the statistics could increase up to four times.

	\section{Tracking efficiency}
	
	The correct estimation of the tracking efficiency is crucial for many analyses focused on determining the branching ratio or production cross-section. It is also a key characteristic in order to qualify tracking algorithms and to find possible implicit sources of inefficiency.
	
	An accessible technique to extract the tracking efficiency consists in using the simulated information as: 
	
	\begin{equation}\label{eq:Eff_MC}
	\epsilon = \frac{N_{reconstructed}}{N_{reconstructible}}
	\end{equation} 
	where $N_{reconstructed}$ is the number of tracks reconstructed by the tracking algorithms and $N_{reconstructible}$ is the number of tracks that fulfils the minimum requirements defined for each subsystem to allow reconstruction.
	However, a method relying solely on simulated information is sensible to a simulation not reproducing perfectly the real data. For this reason, having a method to compute the efficiency using real data is crucial. At LHCb, a tag-and-probe method~\cite{TagProbe} is used exploiting $J/\Psi \to  \mu\mu$ decays to determine the efficiency in data and to apply according corrections to simulation.
	
	In the next section, a novel method to extract the performance of downstream tracking algorithms will be presented, which can be run over both, simulated samples and real data, allowing to validate the method and to extract the correct efficiency respectively.
	
	\subsection{Principle of the method}
	
\begin{figure}[htbp]
		\centering
		\includegraphics[width=0.48\linewidth]{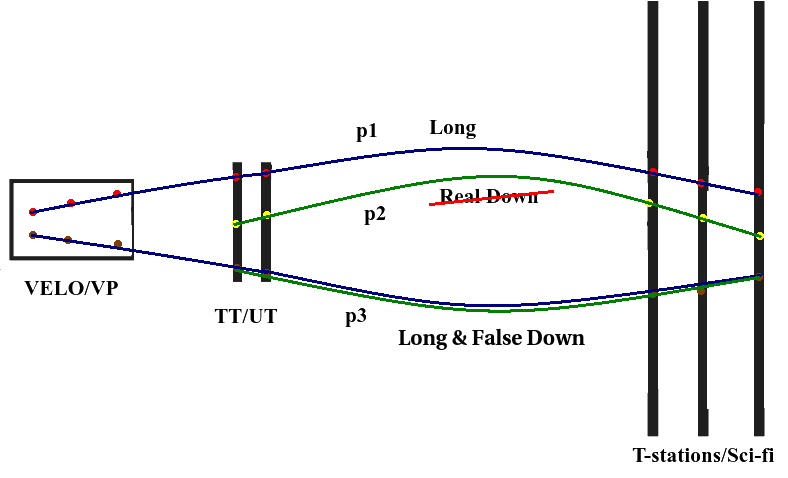}
		\caption{Sketch of the method showing a long track, a real downstream track and a false downstream track associated with a long track.}
		\label{fig:method}
	\end{figure}

	The method uses long tracks reconstructed as downstream tracks from $\Lambda^0\to p\pi$ decays to evaluate the performance of downstream algorithms. 
	The algorithms used to reconstruct both track types are different: long tracks derive from VELO seeds which are extended with hits in the T-stations (forward)~\cite{PatForward}, and the downstream tracks proceed from seeds in the T-stations~\cite{PatSeeding}\cite{Hybrid} which are matched to hits in the TT (backward)~\cite{PatLong}. 
	Technically, this method applies the algorithms for long and downstream tracks reconstruction to all the hits, even those previously used to produce long tracks, producing two different downstream track categories: ``real" and ``false". The ``real" downstream tracks are the ones that would appear in standard condition, i.e. those composed of hits that were not used to reconstruct a long track. The ``false" downstream tracks correspond to actual long tracks, reconstructed by the downstream tracking algorithm without considering the VELO information. The three track categories considered in this method can be seen in Figure~\ref{fig:method}. The ``false" downstream tracks are selected as those which share a high number of hits with a long track in the downstream region (TT and T-stations). Only the ``false" downstream tracks will be considered to estimate the efficiency. 
	
	The efficiency is computed as shown in Equation~\ref{eq:Eff_This}:
	
	\begin{equation}\label{eq:Eff_This}
	\epsilon = \frac{N(\Lambda^0)_{FD}}{N(\Lambda^0)_{L}}
	\end{equation}
	
	where $N(\Lambda^0)_{FD}$ is the number of ``false" downstream tracks (matched to a long track) and $N(\Lambda^0)_{L}$ is the number of long tracks. In both cases the tracks come from a long-lived particle.

	The sketch of the method is shown in Figure~\ref{fig:method} where: first, the long tracking algorithms reconstruct the tracks for particles $p1$ and $p3$ but not $p2$ because it doesn't have any hits in the VELO. Second, the downstream algorithms reconstruct $p2$ and $p3$, since both have hits in TT and T-stations, but not $p1$ due to inefficiencies in the algorithm. Then, the downstream and long tracks for $p3$ are matched. Finally, the efficiency is computed as the number of reconstructed and matched downstream tracks ($p3$) over the number of reconstructed long tracks ($p1$ and $p3$). 
	
	\subsection{Proof of principle}
\begin{figure}[htbp]
		\centering
		\includegraphics[width=0.44\linewidth]{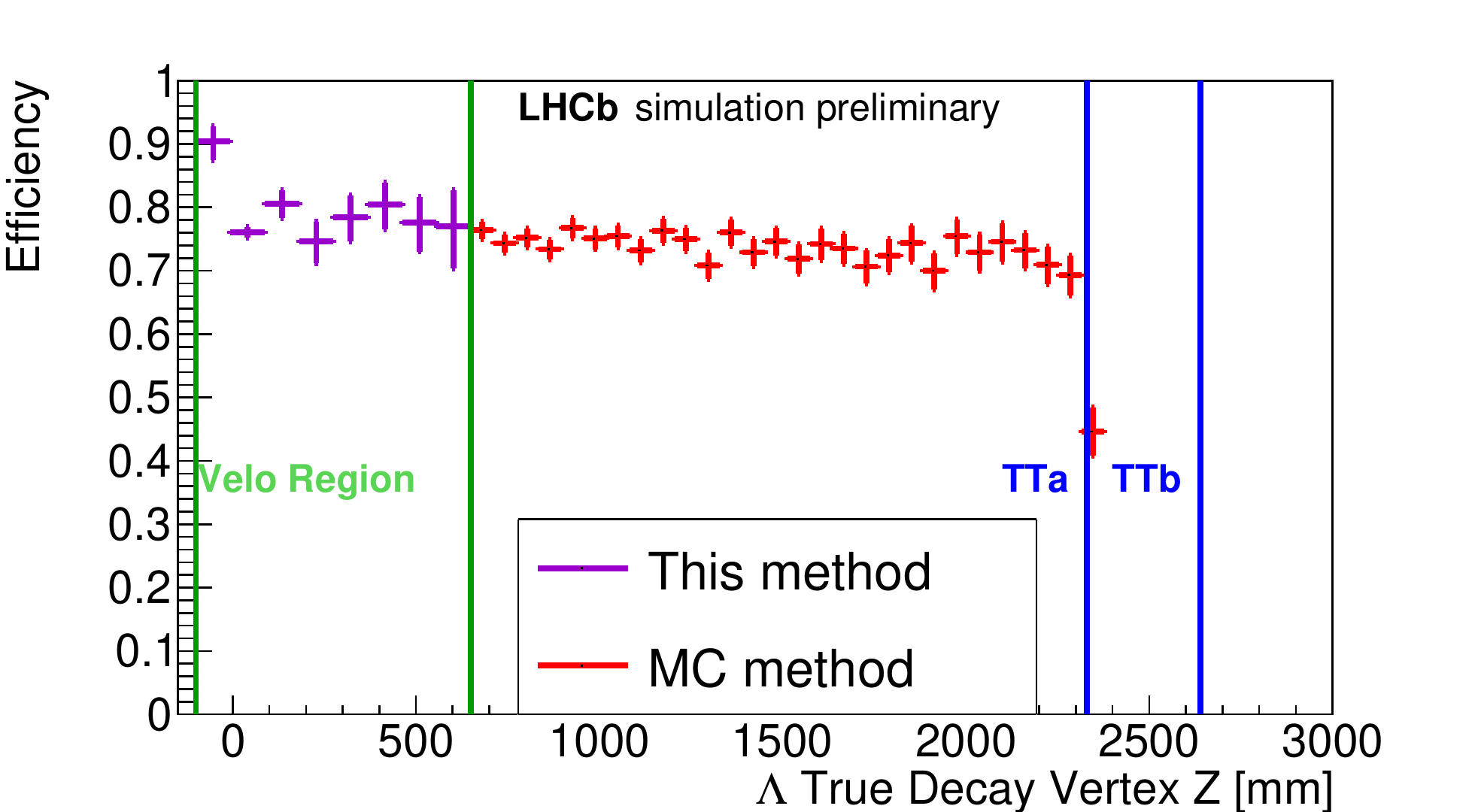}
		\includegraphics[width=0.44\linewidth]{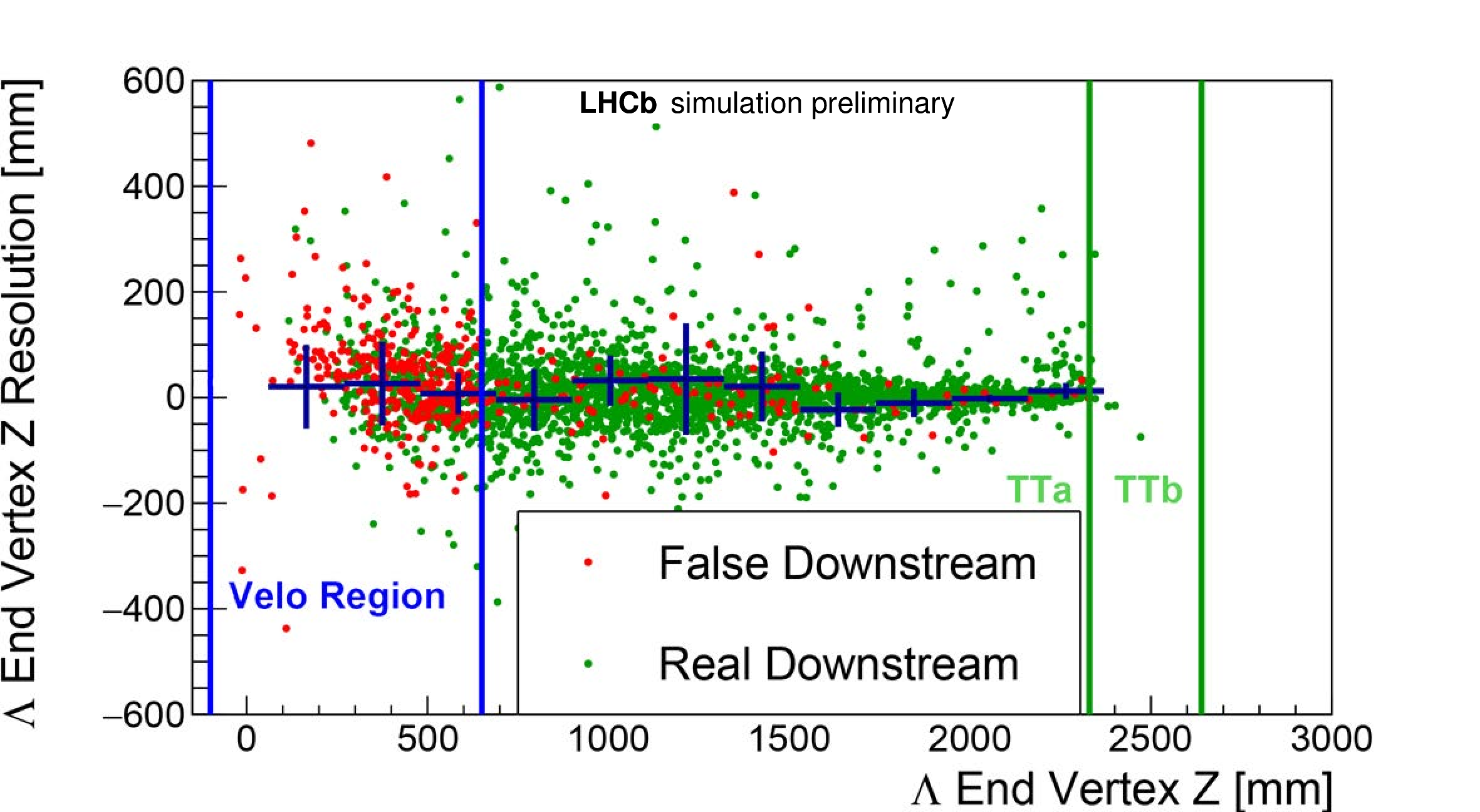}
		\caption{Left: Downstream tracking efficiency as function of the $z$ coordinate of the $\Lambda^0$ decay vertex for the method presented in this article (violet) as compared to the MC method (red). Right: Resolution of the $z$ coordinate of the $\Lambda^0$ decay vertex as function of that $z$ position for ``false" downstream tracks (red) and ``real" downstream tracks (green).}
		\label{fig:proof}
	\end{figure}

	This method relies on the hypothesis that the downstream algorithms can work on the VELO region in the same way as they do it between the VELO and the TT. For this:
	
	\begin{itemize}
		\item The downstream tracking efficiency should not depend on the origin of the track, as shown in Figure~\ref{fig:proof} (Left).
		\item Results on simulation should be compatible with other methods based on truth information from simulation (MC method) as can be seen in Figure~\ref{fig:proof} (Left).
		\item The momentum and spatial resolution from  ``false" downstream tracks should be compatible with the one from ``real" downstream tracks, as shown in Figure~\ref{fig:proof} (Right). No discontinuity or bias is observed.
	\end{itemize}

	\subsection{Results}
	
\begin{figure}[htbp]
		\centering
		\includegraphics[width=0.44\linewidth]{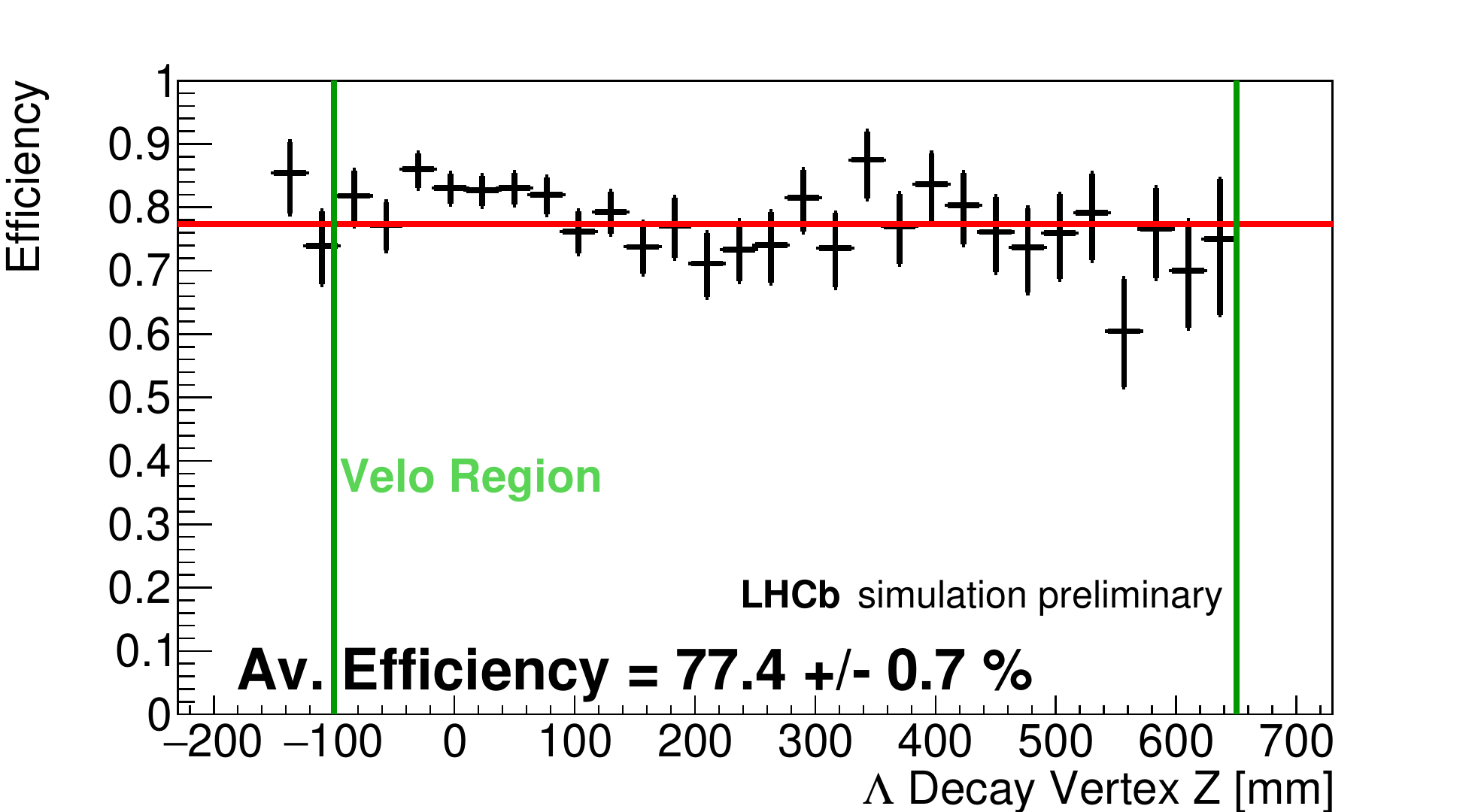}
		\includegraphics[width=0.44\linewidth]{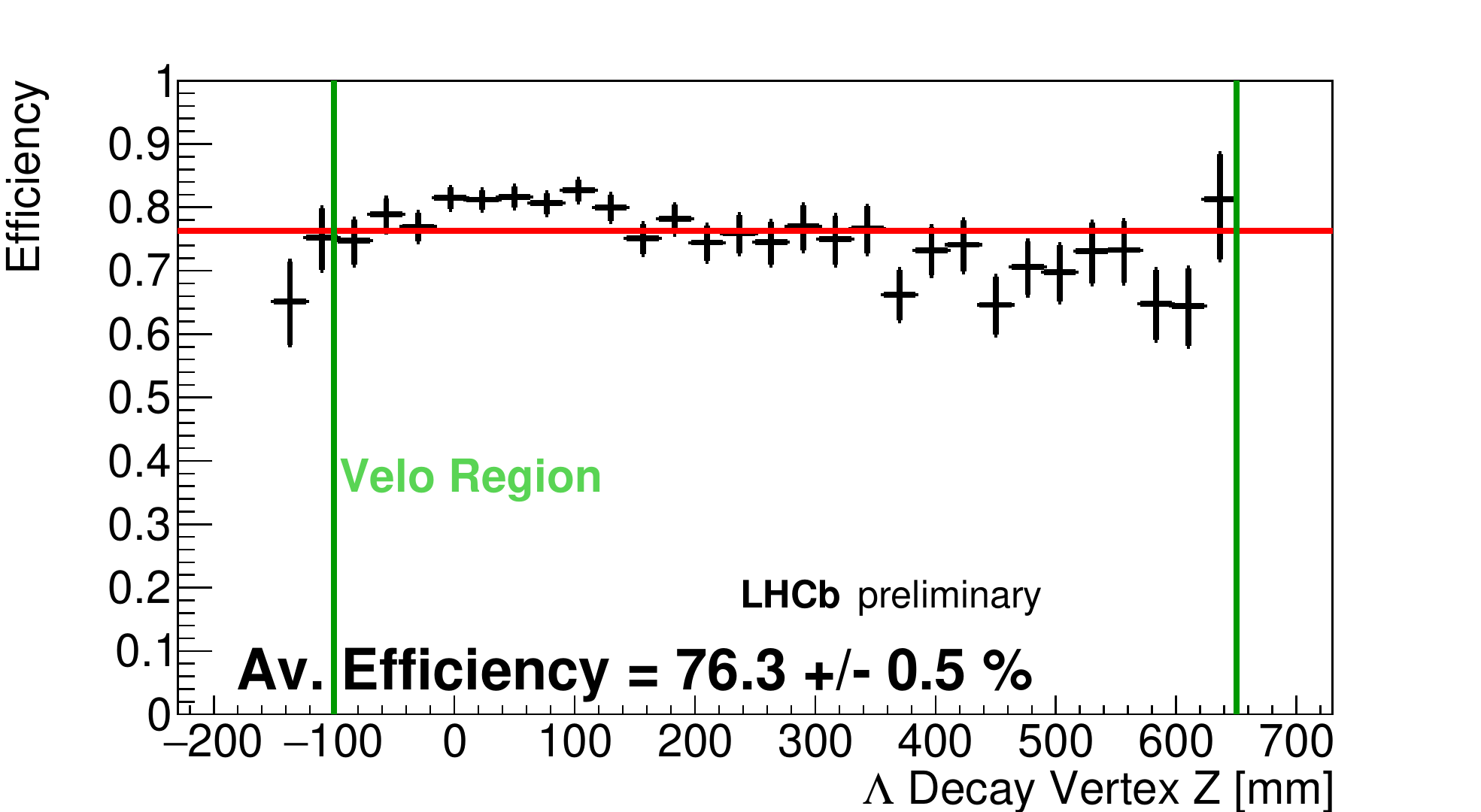}
		
		\caption{Downstream tracking efficiency as function of the $z$ coordinate of the $\Lambda^0$ decay vertex for simulation Run 2 (left), real data Run 2 (right).}
		\label{fig:eff_L0EndVertex}
	\end{figure}

\begin{figure}[htbp]
		\centering
		\includegraphics[width=0.49\linewidth]{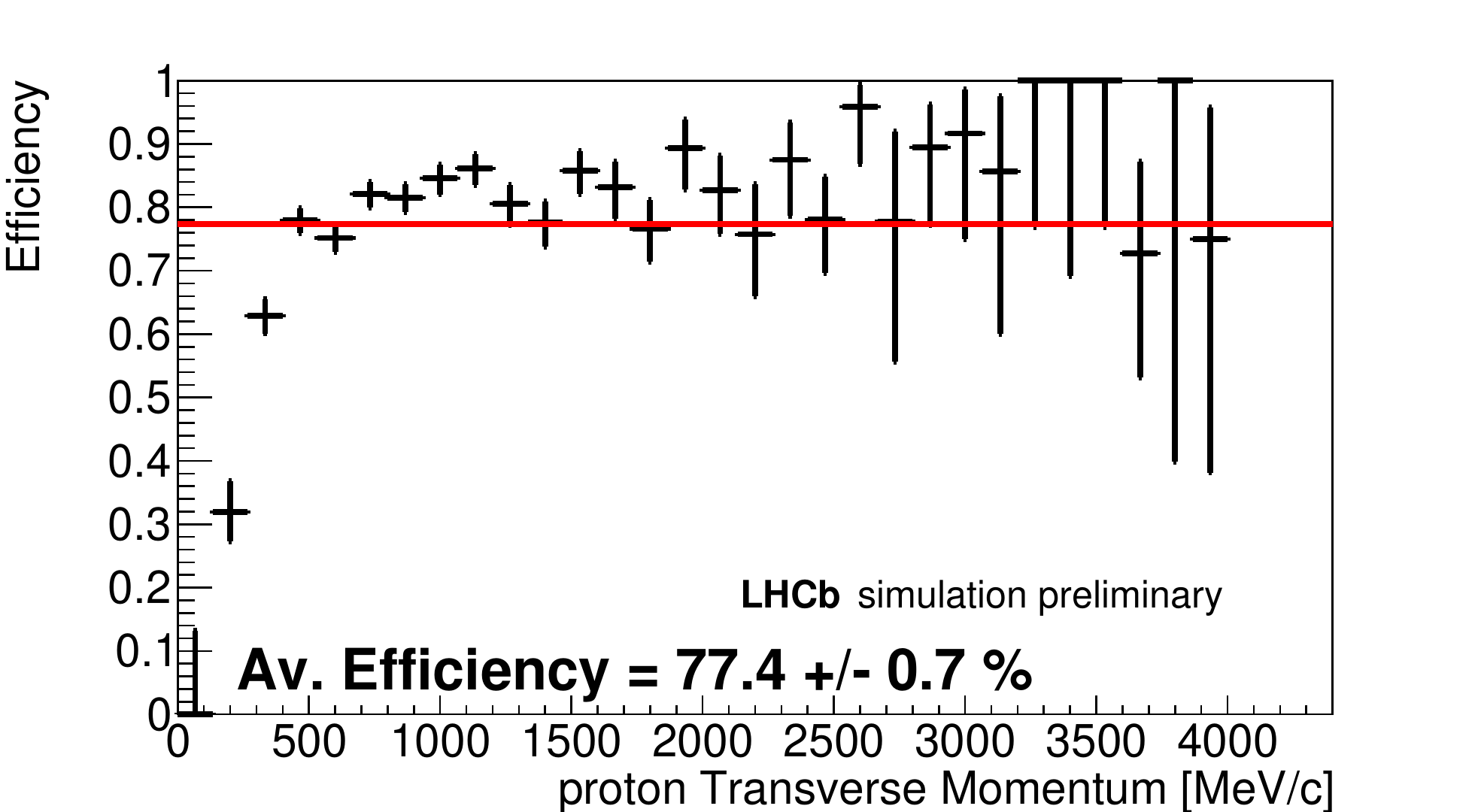}
		\includegraphics[width=0.49\linewidth]{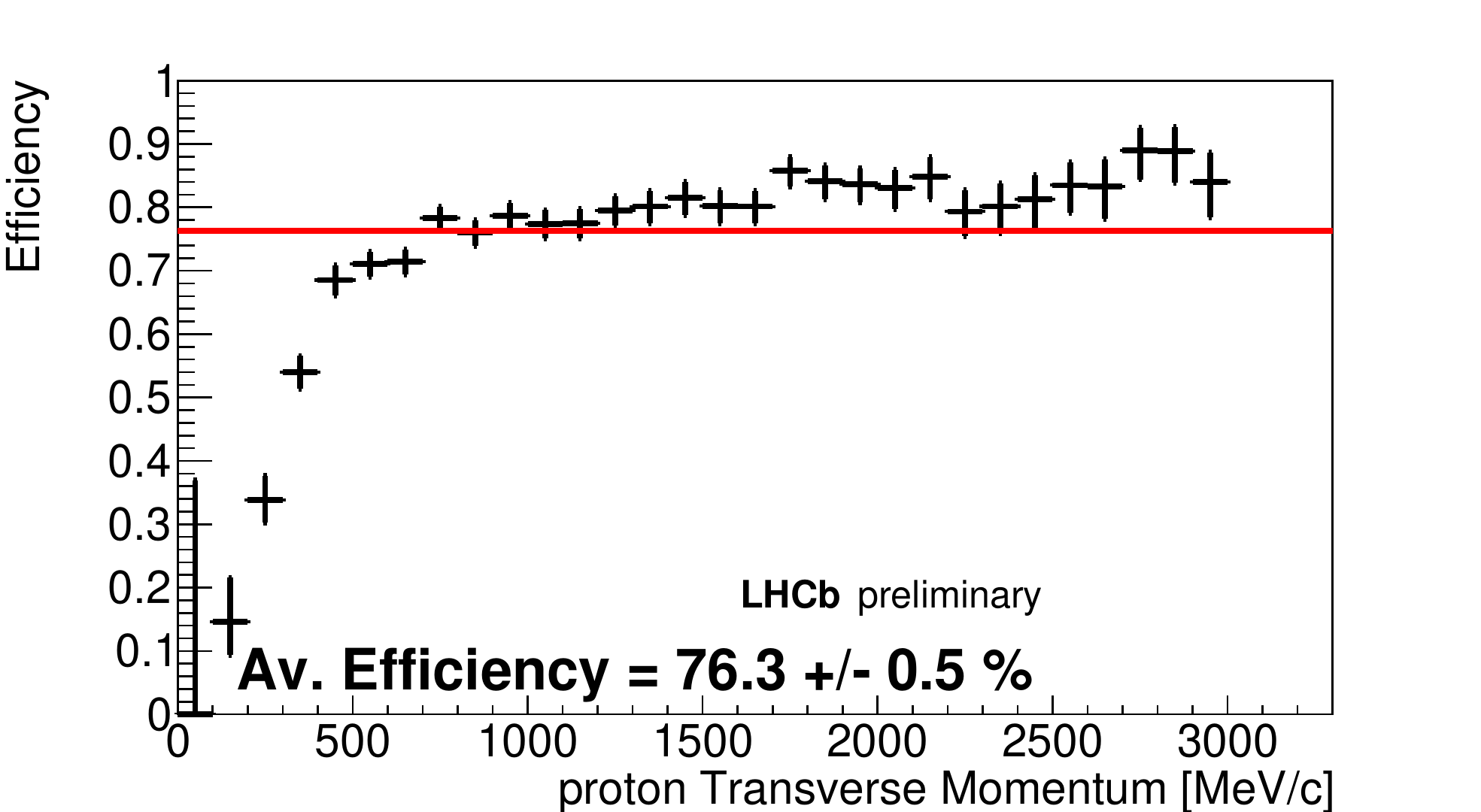}
		\caption{Downstream tracking efficiency as function of the transverse momentum of the track for simulation Run 2 (left), real data Run 2 (right).}
		\label{fig:eff_PT}
	\end{figure}

\begin{figure}
		\centering
		\includegraphics[width=0.49\linewidth]{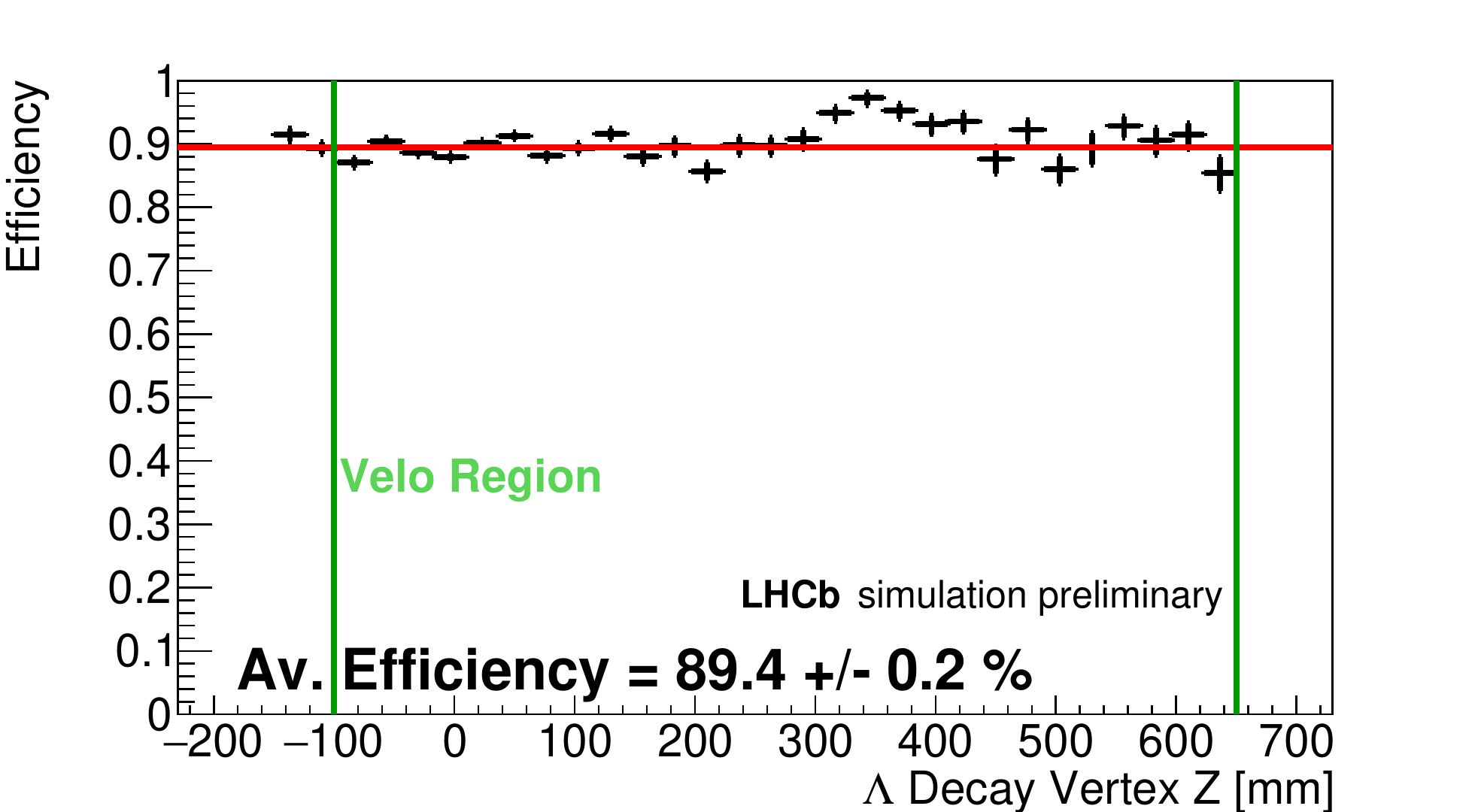}
		\includegraphics[width=0.49\linewidth]{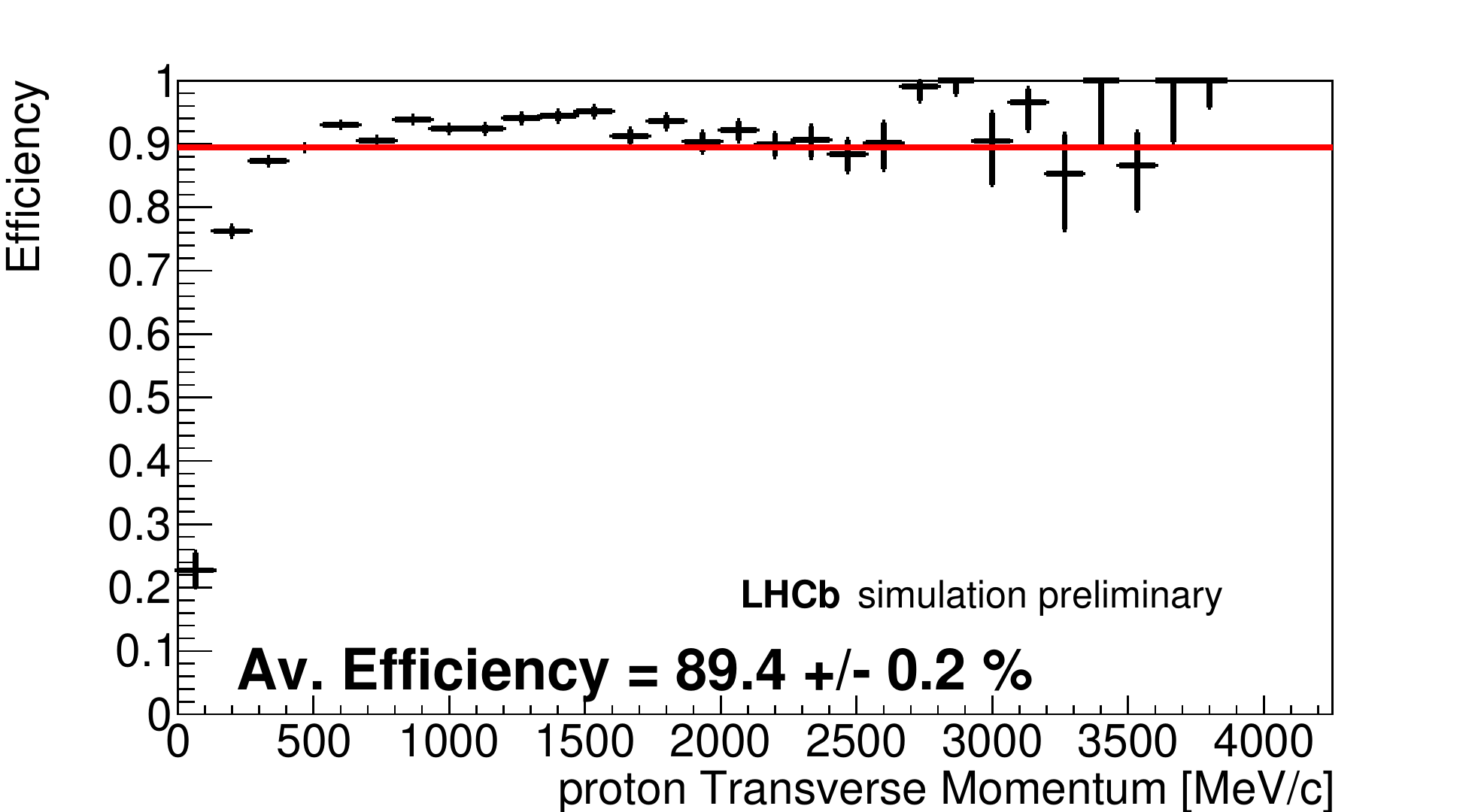}
		\caption{Downstream tracking efficiency as function of the $z$ coordinate of the $\Lambda^0$ decay vertex (left) and the transverse momentum of the track (right) for simulation upgrade conditions.}
		\label{fig:eff_Upgrade}
	\end{figure}

	The method can be used to extract the average value of the downstream tracking efficiency and its dependence on different variables. In Figure~\ref{fig:eff_L0EndVertex}, the efficiency as a function of the $z$ coordinate of the $\Lambda^0$ decay vertex inside the VELO region is shown. It corresponds to the track origin. No dependence of the efficiency with the starting point of the track is observed, both for the simulation and the real data sample. In addition, the efficiency is compatible in the whole range for both cases.
	The downstream tracking efficiency as a function of the transverse momentum of the track is shown in Figure~\ref{fig:eff_PT} for simulation and real data samples. The efficiency is also compatible and shows a similar behaviour.
	This method can also be applied on samples with the upgrade LHCb conditions to test the performance of the tracking algorithms with the new detector. Figure~\ref{fig:eff_Upgrade} shows the efficiency of the downstream tracking algorithms as a function of $z$ coordinate of the $\Lambda^0$ decay vertex and of the transverse momentum of the track for simulation upgrade conditions. 
	Two conclusions can be drawn from this result. On the one hand, the efficiency will continue to be independent of the starting point of the track. On the other hand, the average efficiency is expected to be largely 
	improved thanks to the optimisation of the algorithms and the upgrade of the LHCb detector. In addition, an improvement of the efficiency in the low $p_T$ region will be achieved.
	%
	
	This method assumes that there is no correlation in the reconstruction of long and downstream track. This correlation is expected to be low since the reconstruction algorithms and the track seeds are different. 
	However, any inefficiency related to detector effects are shared among all reconstruction algorithms. Thus further studies will be performed to evaluate potential systematic uncertainties related to this correlation.

	\section{Conclusions}
	
	A new method has been developed to evaluate the performance of downstream tracking algorithms at LHCb using real data as well as simulated one. The results obtained on simulation using this method are in reasonable agreement with other methods. For results based on Run II simulation further studies which will include systematic uncertainties will be performed 
	The performance obtained with this method for downstream tracking algorithms using simulation is compatible with the one obtained using real data. This method can be used to calibrate the downstream tracking algorithms with real data during Run 3. We have tested that the efficiency will largely increase with the upgraded LHCb detector. A summary of the results is shown in Table~\ref{tab:sum_eff}.
	\begin{table}[htbp]	
		\begin{center}
			
			\begin{tabular}{c|c|c|}
				\cline{2-3}
				& \multicolumn{2}{c|}{Efficiency (\%)}     \\ \cline{2-3} 
				& \textbf{This method}    & MC Info        \\ \hline
				\multicolumn{1}{|c|}{Simulation Run II}  & $\mathbf{77.4 \pm 0.7}$ & $74.5 \pm 0.3$ \\ \hline
				\multicolumn{1}{|c|}{Real Data Run II}   & $\mathbf{76.3 \pm 0.5}$ & -             \\ \hline
				\multicolumn{1}{|c|}{Simulation Run III} & $\mathbf{89.4 \pm 0.2}$ & $89.7 \pm 0.1$~\cite{thesis} \\ \hline
			\end{tabular}
		\end{center}
		\caption{Downstream tracking efficiencies for real data and simulation in both, Run 2 and Upgrade conditions, extracted with the method presented at this document. The comparison with the MC method is also presented in the Table. The efficiencies presented here are averaged over the phase space of the tracks of the $\Lambda\to p \pi$ decay.}
		\label{tab:sum_eff}
	\end{table}
	As a final remark, the method presented here can be extrapolated to other track types (Upstream and T-tracks) or to any other experiment with a similar track topology.

\end{document}